\pdfoutput=1
\documentclass[12pt,american,english]{aastex}
\usepackage[T1]{fontenc}
\usepackage[latin9]{inputenc}
\usepackage{array}
\usepackage{amsbsy}
\usepackage{graphicx}

\makeatletter

\providecommand{\tabularnewline}{\\}

\makeatother

\usepackage{babel}
\begin{document}

\title{The asymmetry of sunspot cycles and Waldmeier relations as due to
nonlinear surface-shear shaped dynamo}

\author{V.V. Pipin$^{1-3}$ and A.G. Kosovichev$^{3}$}

\affil{{ $^{1}$ Institute of Geophysics and Planetary Physics, UCLA, Los
Angeles, CA 90065, USA \\
 $^{2}$Institute of Solar-Terrestrial Physics, Russian Academy of
Sciences, \\
 $^{3}$Hansen Experimental Physics Laboratory, Stanford University,
Stanford, CA 94305, USA }}
\begin{abstract}
The paper presents a study of a solar dynamo model operating in the
bulk of the convection zone with the toroidal magnetic field flux
concentrated in the subsurface rotational shear layer. We explore
how this type of dynamo may depend on spatial variations of turbulent
parameters and on the differential rotation near the surface. The
mean-field dynamo model takes into account the evolution of magnetic
helicity and describes its nonlinear feedback on the generation of
large-scale magnetic field by the $\alpha$-effect. We compare the
magnetic cycle characteristics predicted by the model, including the
cycle asymmetry (associated with the growth and decay times) and the
duration - amplitude relation (Waldmeier's effects), with the observed
sunspot cycle properties. We show that the model qualitatively reproduces
the basic properties of the solar cycles.
\end{abstract}

\section{Introduction}

The sunspot's activity is organized on large scales, forming the Maunder
butterfly diagram. It is believed to represent the time - latitude
pattern of the large-scale toroidal magnetic field generated in the
convection zone. Another component of the solar activity is represented
by the global poloidal magnetic field extending outside the Sun and
shaping the solar corona. Both components synchronously evolve as
the solar 11-year cycle progresses. The global poloidal field reverses
the sign in the polar regions near the time of maxima of sunspot
activity. Most of the current solar dynamo models suggest that the
toroidal magnetic field that emerges on the surface and forms sunspots
is generated near the bottom of the convection zone, in the tachocline
or just beneath it in a convection overshoot layer (see, e.g.,
\citealp{1995A&A...296..557R,1995A&A...303L..29C,2007sota.conf..319T,1993ApJ...408..707P}).
The belief in a deep-seated solar dynamo comes from the fact that
this region is sufficiently stable to store magnetic flux despite
the magnetic flux-tube buoyancy effect. However, observations of rotation
rates of emerging magnetic flux within the latitude bands $\pm30^{\circ}$
support a concept of relatively shallow sunspots \citep{1999ApJ...517L.163B}
possibly rooted within the subsurface rotational shear layer. This
concept has support from local helioseismology as well \citep{2011JPhCS.271a2001B}
.

There are further theoretical arguments that the subsurface angular
velocity shear can play an important role in the dynamo process distributed
in the convection zone \citep{2005ApJ...625..539B}. In our previous
paper \citep{pk11apjl} (PK11) we proposed a model of a subsurface-shear
shaped solar $\alpha\Omega$ dynamo. Our model shows that allowing
the large-scale toroidal magnetic field to penetrate into the surface
layers of the Sun changes the direction of the latitudinal migration
of the toroidal field belts and produces the magnetic butterfly diagram
in a good qualitative agreement with the solar cycle observations.
The dynamo wave penetrates close to the surface and propagates along
iso-surfaces of the angular velocity in the subsurface rotational
shear layer in agreement with the Parker-Yoshimura rule \citep{1975ApJ...201..740Y}.
The standard boundary condition typically used in dynamo theories
is to match the internal solution to the potential magnetic field
extending outside of the dynamo region. This boundary condition does
not allow to the toroidal component to penetrate to the surface.

In our previous model the penetration of toroidal magnetic fields
to the surface was modeled by a special boundary condition at the
top of the dynamo region. This boundary condition was formulated as
a linear combination of vacuum (potential field) and perfectly conducting
plasma conditions. The perfectly conducting part results in an increase
of the toroidal component of the large-scale magnetic field at the
boundary. Such boundary condition, used in PK11, models a partial
penetration of the toroidal field into the solar atmosphere, but from
the physical point of view such formulation is rather artificial.
The penetration to the surface can be modeled more physically by
extending the computational domain close to the surface and using
the magnetic diffusivity profile that follows from the standard solar interior
model. This diffusivity decreases toward to the surface and results
in increasing of the toroidal magnetic field in that direction (and
an increase of the gradient of the toroidal magnetic field as well).
The decrease of the turbulent diffusivity, $\eta_{T}\sim{\displaystyle \frac{1}{3}}u'\ell$,
(where $u'$ is the convective RMS velocity and $\ell$ is the mixing
length) is predicted by the mixing-length theory of the solar interior.
This motivates us to extend the integration domain from $(0.71\div0.97)R_{\odot}$,
used in PK11, to $(0.71\div0.99)R_{\odot}$. The convection model
of \citet{stix:02} predicts that towards to the surface the mixing-length, $\ell$, decreases
much faster than $u'$ increases. Figure 1c shows the radial profile
of the turbulent diffusivity in the convection zone model. In this
paper we study how the sharp decrease of the magnetic diffusivity
influences the strength and distribution of the toroidal field in
the upper layers of the convection zone.

There is another reason for extending the integration domain closer
to the surface. In the near surface layers the density stratification
gradient is very strong compared to the bulk of the convection zone.
The mean-field theory predicts a downward turbulent drift of the large-scale
magnetic filed in the presence of the density stratification gradient
\citep{kit:1991} (similar results were obtained by \citealp{pi08Gafd}).
The effective downward drift of large-scale magnetic field results
from magnetic fluctuations in the stratified turbulence. It can be
interpreted as follows (see \citealp{kit:1991}). The intensity of
the magnetic fluctuations $\overline{b^{2}}=\mu_{0}\bar{\rho}\overline{u'^{2}}$
($\overline{\rho}$ is the mean density) rises in the direction of
the density gradient because the turbulent RMS velocity varies slower
than $\overline{\rho}$. Random Lorenz forces, which are induced by
small-scale magnetic fields $\mathbf{b}$ and large-scale field $\overline{\mathbf{B}}$,
produce fluctuating flows $\mathbf{u}'\approx{\displaystyle \frac{\mathbf{\left(\nabla\times\mathbf{b}\right)}\times\overline{\mathbf{B}}}{\mu\overline{\rho}}\tau_{c}}$.
The resulted electromotive force $\overline{\mathbf{u}'\times\mathbf{b}}$
is perpendicular to the large-scale field. { This can be interpreted as
an effective downward velocity drift of the large-scale magnetic
field \citep{kit:1991}.}
The theory also predicts that this
kind of turbulent pumping is quenched by the influence of the Coriolis
force, which results in the velocity of the effective drift to be
greatest near the surface where the density stratification is strong.
Thus, qualitatively, this effect works similarly to so-called {}``topological
pumping'' \citep{1974JFM....65...33D} .

Our study includes an equation of the magnetic helicity evolution,
proposed by \citet{kleruz82} and \citet{kle-rog99}. This equation
describes the balance between the small-scale turbulent magnetic helicity
and the large-scale magnetic helicity generated by the dynamo process,
and has been used in many previous dynamo studies (e.g., \citet{2005PhR...417....1B}
and references therein). One of our goals is to explore nonlinear
feedback of the magnetic helicity on the basic properties of sunspot
cycles, e.g., the relationship between the rise and decay times, and
between the length and strength of the cycles. The results for a dynamo
model in a single-mode approximation \citep{2009GApFD.103...53K,2010IAUS..264..202K}
have suggested the importance of the nonlinear magnetic helicity effects
for the solar-cycle behavior. The next section describes the formulation
of the 2D mean-field dynamo model, including the basic assumptions, the
reference model of the solar convection zone, and input parameters
of the large-scale flows. Section 3 presents the results and discussion.
The main findings are discussed in Section 4.

\section{Basic equations}

The dynamo model is based on the standard mean-field induction equation
in perfectly conductive media (Krause and Rädler, 1980):
\[
\frac{\partial\mathbf{B}}{\partial t}=\boldsymbol{\nabla}\times\left(\mathbf{\boldsymbol{\mathcal{E}}+}\mathbf{U}\times\mathbf{B}\right)
\]
 where $\boldsymbol{\mathcal{E}}=\overline{\mathbf{u\times b}}$ is
the mean electromotive force, with $\mathbf{u,\, b}$ being the turbulent
fluctuating velocity and magnetic field respectively; $\mathbf{U}$
is the mean velocity. General expression for $\boldsymbol{\mathcal{E}}$
was obtained by \citet{pi08Gafd} (hereafter P08). Following Krause
and Rädler(1980) we write the expression for the mean electromotive
force as follows:
\begin{equation}
\mathcal{E}_{i}=\left(\alpha_{ij}+\gamma_{ij}\right)\overline{B}_j-\eta_{ijk}\nabla_{j}\overline{B}_{k}.\label{eq:EMF-1}
\end{equation}
Tensor $\alpha_{i,j}$ represents the alpha effect, including the
hydrodynamic and magnetic helicity contributions, $\alpha_{ij}=C_{\alpha}\psi_{\alpha}(\beta)\sin^{2}\theta\alpha_{ij}^{(H)}+\alpha_{ij}^{(M)},$
where the hydrodynamical part of the $\alpha$-effect, $\alpha_{ij}^{(H)}$,
and the quenching function, $\psi_{\alpha}$, are given in Appendix
(see also in \citealp{pk11apjl}), the parameter $C_{\alpha}$ controls
the amplitude of the $\alpha$-effect. The hydrodynamic $\alpha$-effect
term is multiplied by $\sin^{2}\theta$ ($\theta$ is colatitude)
to prevent the turbulent generation of magnetic field at the poles.
The contribution of the small-scale magnetic helicity $\overline{\chi}=\overline{\mathbf{a\cdot}\mathbf{b}}$
($\mathbf{a}$ is a fluctuating vector-potential of magnetic field)
to the $\alpha$-effect is defined as $\alpha_{ij}^{(M)}=C_{ij}^{(\chi)}\overline{\chi}$
, where coefficient $C_{ij}^{(\chi)}$ depends on the turbulent properties
and rotation, and is given in Appendix. The other parts of Eq.(\ref{eq:EMF-1})
represent the effects of turbulent pumping, $\gamma_{ij}$, and turbulent
diffusion, $\eta_{ijk}$. They are the same as in PK11. We describe
them in Appendix.

The nonlinear feedback of the large-scale magnetic field to the $\alpha$-effect
is described as a combination of an \textquotedbl{}algebraic\textquotedbl{}
quenching by function $\psi_{\alpha}\left(\beta\right)$ (see Appendix
and PK11), and a dynamical quenching due to the magnetic helicity
conservation constraint. The magnetic helicity, $\overline{\chi}$
, subject to a conservation law, is described by the following anzatz
\citep{kleruz82,kle-rog99}:
\begin{eqnarray}
\frac{\partial\overline{\chi}}{\partial t} & = & -2\left(\boldsymbol{\mathcal{E}\cdot}\overline{\mathbf{B}}\right)-\frac{\overline{\chi}}{R_{\chi}\tau_{c}},\label{eq:hel}
\end{eqnarray}
 where $\tau_{c}$ is a typical convection turnover time. Parameter
$R_{\chi}$ controls the helicity dissipation rate without specifying
the nature of the loss. { Generally, we can expect that the formulation of
the the helicity loss term in Eq.(\ref{eq:hel}) affects properties of
the dynamo solutions. This is suggested by results
that can be found in the literature
\citep{
2007AN....328.1118B,2010AN....331..130M,
2010MNRAS.409.1619G,2011A&A...526A.138M}.  The physics of helicity
loss is poorly understood, and influence the various processes of the
helicity flux loss on the properties of magnetic cycles deserves a separate
study. To reduce the number of free parameters in the model
we consider the simplest form of helicity
flux loss. The  parameter $R_{\chi}$ controls the amount of
the magnetic flux generated by the dynamo. This amount can be roughly
estimated from observations. We use the range of  $R_{\chi}$ that
gives the total magnetic flux of the order of
$\approx 10^{24}-10^{25}$ Mx  in agreement with observations
\citep{1994SoPh..150....1S}.
Another parameter controlling the helicity
dissipation in our model is  $\tau_{c}$. It is given by the solar interior model.}
It seems to be reasonable that the helicity
dissipation is most efficient in the near surface layers because of
the strong decrease of $\tau_{c}$ (see Figure 1b).

We use the solar
convection zone model computed by \citet{stix:02}, in which the mixing-length
is defined as $\ell=\alpha_{MLT}\left|\Lambda^{(p)}\right|^{-1}$,
where $\mathbf{\boldsymbol{\Lambda}}^{(p)}=\boldsymbol{\nabla}\log\overline{p}\,$
is the pressure variation scale, and $\alpha_{MLT}=2$. The turbulent
diffusivity is parametrized in the form, $\eta_{T}=C_{\eta}\eta_{T}^{(0)}$,
where $\eta_{T}^{(0)}={\displaystyle \frac{u'\ell}{3}}$ is the characteristic
mixing-length turbulent diffusivity, $\ell$ and $u'$ are the typical
correlation length and RMS convective velocity of turbulent flows,
respectively and $C_{\eta}$ is a constant to control the intensity
of turbulent mixing. In the paper we use $C_{\eta}=0.05$. The differential
rotation profile, $\Omega=\Omega_{0}f_{\Omega}\left(x,\mu\right)$,
$x=r/R_{\odot}$, $\mu=\cos\theta$ is a modified version of an analytical
approximation to helioseismology data, proposed by \citet{1998MNRAS.298..543A},
see Figure 1a.

We use the standard boundary conditions to match the potential field
outside and the perfect conductivity at the bottom boundary. As discussed
above, the penetration of the toroidal magnetic field in to the near
surface layers is controlled by the turbulent diffusivity and pumping
effect (see Figures 1c and 1d).

\begin{table}
\begin{center}
\begin{tabular}{|>{\centering}m{1cm}|>{\centering}m{1.5cm}|>{\centering}m{1cm}|>{\centering}m{1.5cm}|>{\centering}m{1cm}|>{\centering}m{1.9cm}|>{\centering}m{1cm}|>{\centering}m{1.5cm}|>{\centering}m{2cm}|}
\hline
Model  & $C_{\alpha}$ & $R_{\chi}$ & ${\displaystyle \frac{\left.\eta_{T}\right|_{0.99R_{\odot}}}{\max(\eta_{T})}}$ & $r\boldsymbol{\Lambda}_{min}^{(\rho)}$ & $\left\langle \! B_{\phi}\!\right\rangle _{SL}$,

G & $B_{r}^{Polar}$, G & $A_{W}$ & {\it Period},

Yr\tabularnewline
\hline
P1

P2

P3 & 0.03 & - & 0.245  & -160

-30

0 & 1000

450

60 & 7.2

3.5

0.15 & 0.74

0.94

0.89 & 13.4

11.4

8.6\tabularnewline
\hline
D1

D2 & 0.03 & - & 0.645

0.091 & -160 & 450

1600 & 3.6

14.4 & 0.78

0.68 & 10.7

15.2\tabularnewline
\hline
CQ1

CQ2

CQ3 & 0.03 & $5\!\cdot\!10^{4}$

$10^{3}$

$10^{2}$ & 0.245 & -160 & -

66

152 & -

0.3

3.6 & -

0.71

0.59 & -

11.74

11.17\tabularnewline
\hline
WR1 & 0.03,0.04

0.05,0.06 & 50 & 0.245 & -160 & 203,294

351,396 & 1.1,1.8

2.3,2.6 & 0.68,0.54

0.5,0.42 & 11.35,10.60

10.15,9.67\tabularnewline
\hline
WR2 & 0.03,0.04

0.05,0.06 & 100 & 0.245 & -160 & 152,220

266,302 & 0.8,1.3

1.8,2.0 & 0.59,0.53

0.49,0.39 & 11.17,10.44

9.80,9.33\tabularnewline
\hline
WR3 & 0.03,0.04

0.05,0.06 & 200 & 0.245 & -160 & 102,150

182,206 & 0.5,0.9

1.2,1.4 & 0.57,0.50

0.44,0.38 & 11.10,10.40

9.80,9.27\tabularnewline
\hline
\end{tabular}
\par\end{center}

\caption{Parameters and characteristics of the dynamo models: the $\alpha$-effect
parameter $C_{\alpha}$; parameter $R_{\chi}$ controls the helicity
dissipation rate; ${\displaystyle \frac{\left.\eta_{T}\right|_{0.99R_{\odot}}}{\max(\eta_{T})}}$
is the ratio between the maximum background turbulent diffusivity
and the value at the top boundary (in the reference model of Stix
(2002) this value is 0.245); $r\boldsymbol{\Lambda}_{min}^{(\rho)}$
is the minimum of the density gradient height at the top; $\left\langle B_{\phi}\right\rangle _{SL}$
is the maximum strength of the toroidal magnetic field averaged in
the range of $0.95-0.99R_{\odot}$; $B_{r}^{Polar}$ is the maximum
strength of the radial polar magnetic field at $0.99R_{\odot}$; $C_{W}$
is a calibration coefficient used for calculation of the sunspot number
parameter; $A_{W}$ is the sunspot number asymmetry parameter obtained
in the models; {\it Period} is the period of sunspot cycles from the dynamo
models.}
\end{table}

\section{Results}

We summarize the parameters and characteristics of the dynamo models
in Table 1. For simulating the sunspot number obtained from solar
observations we use the relation suggested by \citet{1988MNRAS.230..535B}:
\begin{equation}
W(t)=C_{W}\,\max\left(\left|\left\langle B(r,\theta,t)\right\rangle _{0.95-0.99R}\right|,0<\theta<180\right)^{3/2}\label{eq:Wt}
\end{equation}
where $B$ is the toroidal magnetic field strength and $C_{W}$ is
the calibration coefficient, we use $C_{W}=1/40$. We define the asymmetry
parameter of the cycle as the ratio between the modulus of the mean
decay rate and mean rise rate, $A_{W}={\displaystyle \frac{\overline{\partial_{t}W}\biggr|_{\partial_{t}W>0}}{\left|\overline{\partial_{t}W}\right|\biggr|_{\partial_{t}W<0}}}$.
The simulations were started from initial states with zero toroidal
magnetic field and weak poloidal field which is symmetric about the equator.
The solution is found by a semi-implicit method using a finite-difference
approximation in radius and a pseudospectral decompisition in terms
of Legendre polynomials in latitude. The numerical scheme conserves
the parity of solution with respect to the equator. The characteristics of the dynamo
models were determined from the stationary periodic solutions.

\subsection{Effects of the near surface diffusion and turbulent pumping }

In this part of the paper we fix the $\alpha$-effect parameter $C_{\alpha}=0.03$
(the dynamo instability threshold is $C_{\alpha}\approx0.02$ ). In
the near surface layers the Coriolis number is rather small. \label{The-turbulent-pumping}
Therefore, the turbulent pumping primarily depends on the density
gradient $\boldsymbol{\Lambda}^{(\rho)}$ and diffusion coefficient
$\eta_{T}$. The gradient parameter $r\boldsymbol{\Lambda}^{(\rho)}$
varies from $\approx-7$ at $r=0.71R_{\odot}$ to $\approx-160$ at
$r=0.99R_{\odot}$. To illustrate the influence of the pumping effect
on the dynamo model solution we examine three different cases (P1,
P2 and P3 in Table 1). Model P1 employs the density gradient profile
provided by the \citet{stix:02} model. In model P2, we introduce
an artificial limit on the level of $r\Lambda^{(\rho)}=-30$ suggested
by \citet{2000A&A...359..531K}. In model P3, we completely neglect
the pumping effect. Profiles of the radial and the latitudinal pumping
velocities at $\theta=45^{\circ}$ for models P1 and P2 are shown
in Figure 1d. Note, that compared to the plotted values the amplitude
of the velocities in the models is reduced by factor $C_{\eta}=0.05$.

The time - latitude toroidal magnetic field ``butterfly'' diagrams, which
were averaged over the depths from $r=0.95R_{\odot}$ to $r=0.99R_{\odot}$,
and the radial magnetic field evolution at $r=0.99R_{\odot}$ for
models P1, P2 and P3 are shown in Figures 2. We find that the larger
amplitude of the downward turbulent pumping results in the greater
strength of the near surface toroidal magnetic field. The turbulent
pumping increases the efficiency of the subsurface shear generation
effect. This leads to a faster migration rate of the toroidal magnetic
field to the equator (see, \citealp{1975ApJ...201..740Y}).

A similar effect can be produced by changing the turbulent diffusivity
profile near the surface. In Figure 3 we show the results for two
cases of the increased (D1) and decreased (D2) turbulent diffusivity
(Figure 1c). The diagrams for these two cases can be compared
with the reference Stix's model P1 in Figure 2a. The smaller the surface
turbulent diffusivity level, the greater toroidal magnetic field strength.
In the reference case the typical magnetic field strength is $\sim1$
kG. If the surface turbulent diffusivity is three time smaller the
field strength increases to $\sim1.6$ kG; if the diffusivity is two
time larger, then the field strength is only about $\sim450$ G. The
inclination of the toroidal field patterns migrating toward the equator (latitudinal migration
speed) does not change considerably with the changes of the turbulent
diffusivity profile. However, when the surface turbulent diffusivity
is smaller, the toroidal magnetic field migrates closer to the equator,
and the cycle becomes longer.

For simulating the sunspot number obtained from solar observations
we use the relation motivated by \citet{1988MNRAS.230..535B}, see
Eq.(\ref{eq:Wt}). The estimated sunspot number for the models with
the increased and decreased sub-surface turbulent diffusivity is shown
in Figure 3 (left panels). The asymmetry between the rise and decay
phases is clearly seen for the model with the decreased surface turbulent
diffusivity. In the mean-field dynamo concept the decay phase of the
large-scale toroidal magnetic field is defined by turbulent diffusion
\citep{park}. The decrease of the surface turbulent diffusivity increases
the decay time when the toroidal field is located closer to the surface.

\subsection{Magnetic helicity effect and the Waldmeier's relations }

The evolution of magnetic helicity based on the conservation law is
described by Eq.(\ref{eq:hel}). Without helicity fluxes from the
dynamo domain \citep{2000A&A...361L...5K,vish-ch:01} or in absence
of helicity dissipation, the generation of magnetic helicity by dynamo
leads to {}``catastrophic quenching'' of the $\alpha$-effect, which
stops the dynamo process (see, e.g.,
\citealp{1992ApJ...393..165V,2000A&A...361L...5K,2005PhR...417....1B}).
In our model the dissipation of magnetic helicity is described by
a decay term:${\displaystyle -\frac{\overline{\chi}}{R_{\chi}\tau_{c}}}$.
We illustrate the catastrophic quenching in dynamo model CQ1 in Figure
4, which shows the results for $R_{\chi}=5\times10^{4}$. In this
case, the rate of the helicity loss from the Sun is small, and the
dynamo process stops after 3-4 periods (2 magnetic cycles). The time-latitude
diagram for the current helicity, which is estimated as
 $h_{\mathcal{C}}=\overline{\mathbf{b\cdot}\boldsymbol{\nabla}\times\mathbf{b}}\approx\overline{\chi}/\ell^{2}$,
is shown together with the magnetic butterfly diagram. We see that
the total magnetic helicity generated by the dynamo is decaying much
slower than the dynamo waves. Figure 4 also shows the evolution of
the sunspot parameter, the total magnetic flux, the total turbulent
magnetic helicity and total large-scale magnetic helicity. For comparison
with case $R_{\chi}=5\times10^{4}$ (model CQ1), we show the results
for $R_{\chi}=10^{3},\,10^{2}$ (models CQ2 and CQ3). In the case
of $R_{\chi}=10^{3}$ (model CQ2) the dynamo is stabilized at a quite
low level with the maximum toroidal magnetic field strength of about
$50\,\mathrm{G}$ and the total magnetic flux of about $10^{24}$
Mx. For the high dissipation rate of the magnetic helicity, $R_{\chi}=10^{2}$,
(model CQ3) the maximum of the toroidal magnetic field strength inside
the convection zone is about $450\,\mathrm{G}$, and in the surface
layer it reaches about $150\,\mathrm{G}$. In model CQ3, the total
magnetic flux is about $4\times10^{24}$ Mx. This roughly agrees with
observational results of \citet{1994SoPh..150....1S}. Taking into
account the results shown in Figure 4d we can estimate the amount
of the helicity loss from the Sun per cycle. In the model CQ3 it is
about $1.26\times10^{45}\,\mathrm{Mx^{2}}$.

Model CQ3 is most relevant for comparison with observations. This
case qualitatively reproduces the basic features of the solar cycle.
Figure 5 shows snapshots of toroidal and poloidal fields,
the time-latitude diagram of the near surface toroidal
magnetic field, and the current helicity evolution, and the
time-radius diagrams for the magnetic field and current helicity
for this model.
{ The time-laltitude diagrams illustrate the migration
of the toroidal and poloidal fields and polarity reversal. The
time-radius diagrams show migration of the magnetic field with radius
at $30^{\circ}$ latitude, and an interesting concentration of the
field at $r/R_{\odot}\sim 0.9-0.92$, or $60-70$Mm below the
surface. This concentration is related to the second
maximum of the dynamo wave when it propogates from the bottom of the
convection zone to the surface.

 Our results show that the current helicity
changes the sign in the near surface layers at the begining of the
cycle. A similar behaviour was found in observations
of \citet{2010MNRAS.402L..30Z} (see their Fig.2). Our initial comparison
of the current helicity pattern produced by the model reveals some
disagreements with observations, e.g., the model does not show
the change of the current helicity
 sign near the equator at the end of sunspot cycle. This problem needs a separate
 study, which should include a more sophisticated description of the
 helicity fluxes.
The time-radius diagram at latitude $30^{\circ}$ for the
toroidal magnetic field and the current helicity (Figure 5e) shows that
that in the bulk of the convection zone the  current helicity
does not change much with time, and  that the helicity is nearly constant near the
bottom of the convection zone. In the north hemisphere the magnetic helicity
is positive at the bottom of the convection zone because the kinetic
part of the $\alpha$ effect is negative there (also see,
\cite{pk11apja} and Figure 1 there)}. In the near-surface layer the sign of the
current helicity changes at the rising phase of the cycle.

Model CQ3 clearly shows asymmetry between the growth and decay phases
of the sunspot number parameter $W(t)$. We find that the asymmetry
increases with increase of the amplitude of the cycle, i.e., with
increase of $\alpha$-effect parameter $C_{\alpha}$ (see Figure 6).
It is expected that for the higher $C_{\alpha}$ the dynamo period
is shorter (see, e.g., \citealp{park}) . This motivates us to look
at the period - amplitude relationship for our model and also at the
amplitude dependence on the growth/decay rate by computing a series
of models WR1, WR2, WR3 for various values of $C_{\alpha}$ and $R_{\chi}$.
The results for the asymmetry parameter as well as the cycle period
are summarized in Table 1. We compare the model results with the asymmetry
estimated from the monthly smoothed sunspot number provided by the
\citeauthor{sidc}. The data set was additionally smoothed by means
of the Wiener filter. After this, we divided the whole data set covering
the time period from 1749 to 2010, into separate sunspot cycles. The
cycles were divided by a program that catches the sequences of the
sunspot minima. For each cycle we estimate the growth rate by a ratio
of the cycle amplitude to the growth time. Similarly, the decay rate
was defined.

Figure 7 compares the model with these estimates in the form of the
Waldmeier's (1935) relations: a) amplitude - rise rate, b) amplitude
- decay rate, c) period - amplitude, d) rise time - amplitude, e)
rise vs decay rates and g) rise vs decay times. The results obtained
from the experimental data set confirm the findings of other authors
(see, \citealp{vit-kop-kuk,2002SoPh..211..357H,2007ApJ...659..801C,2009GApFD.103...53K,2011MNRAS.410.1503K}).
The computed dynamo models (WR1, WR2 and WR3) for a given range of
the $\alpha$-effect parameter $C_{\alpha}=0.03-0.06$ and magnetic
helicity dissipation rate $R_{\chi}=50-200$) correspond reasonably
well to the data points. However, there are differences. One possible
source of the difference between the model and the data is clarified
in Figure 7(e), which shows correlation between the rise and decay
rates in the solar cycles. We find that for most solar cycles the
rise rate is higher than the decay rate. The mean asymmetry parameter
is $\overline{A}_{W}\approx0.68\pm0.31$ . As seen in Figure 7, our
models have smaller $A_{W}$ . Thus, our dynamo models produce more
asymmetric $W$ profiles than the observed sunspot number. Increasing
or decreasing the helicity dissipation by changing $R_{\chi}$ did
not improve the agreement with the observations.

It is clear that more studies of the turbulent properties of the Sun
are necessarily. Nevertheless, the initial results of the dynamo models
shaped by the subsurface shear layer are encouraging.

\section{Discussion and conclusion}

The paper presents a study of a solar dynamo model operating in the
bulk of the convection zone with the toroidal magnetic field flux
shaped into the time-latitude ``butterfly'' diagram
in the subsurface rotational shear layer. We explore
how this type of dynamo may depend on the radial variations of turbulent
parameters and the differential rotation near the surface. The mean-field
dynamo model takes into account the evolution of the magnetic helicity
and describes its nonlinear feedback on the generation of the large-scale
magnetic field by the $\alpha$-effect. We compare the magnetic cycle
characteristics predicted by the model, including the cycle asymmetry
and the duration - amplitude relation (Waldmeier's effects) with the
observed sunspot cycle properties. We show that the model qualitatively
reproduces the basic properties of the solar cycles. However, the
model cycles are systematically more asymmetric than the observed
cycles.

In section 3.1, it was shown that the radial profiles of the turbulent
diffusivity and the density stratification scale in the sub-surface
layer (between $0.95-0.99R_{\odot})$ may significantly influence
the dynamo properties. In particular, the surface-shear shaped dynamo
model favors  the negative gradient of turbulent diffusivity in the
sub-surface layer as follows from the standard solar model. This is
contrary to the positive gradient of turbulent diffusivity often used
in flux-transport dynamo models (e.g., \citealp{2011MNRAS.410.1503K}).
In our model, a steeper gradient of the magnetic diffusivity results
in a stronger toroidal magnetic field, a higher latitudinal migration
speed and a longer magnetic field decay time in the surface layer.
We found that the downward turbulent pumping of the horizontal magnetic
field (associated with either toroidal or meridional magnetic field
components) brings the dynamo properties in better agreement with
observations, increasing the period of the magnetic cycle for a given
turbulent diffusivity profile. The model shows the asymmetry between
the rise and decay rates (and duration phases) of the toroidal magnetic
field. The asymmetry increases with the increase of the turbulent
diffusivity gradient in the sub-surface layer.

The models shows a clear dependence of the asymmetry parameter (the
ratio between the cycle's decay and rise rates) on the magnetic cycle
strength. We compared a sunspot number parameter previously suggested
by \citet{1988MNRAS.230..535B} with statistical properties of the
solar cycle. Our model qualitatively reproduces the known properties,
such as the Waldmeier's relations and the period - amplitude dependence.
In particular, Figure 7e shows that the asymmetry is one of the basic
features of the sunspot cycle activity (see also, \citealp{vit-kop-kuk}).

In our model this asymmetry depends on the parameters of the turbulent
diffusivity in the near surface layer and on the rate of the magnetic
helicity dissipation. If the magnetic helicity dissipation rate is
higher, the asymmetry is smaller. The magnetic helicity dissipation
rate influences the amount of the total magnetic flux produced in
the Sun. According to \citet{1994SoPh..150....1S} the total magnetic
flux produced during a solar cycle is about $10^{24}$$\mathrm{{Mx}}$.
The models presented in the paper satisfy this constraint. The estimated
amount of magnetic helicity loss in the dynamo model is about $10^{45}$
$\mathrm{Mx}^{2}$ per cycle.

Thus, we conclude that the dynamo models with the subsurface shear
layer can satisfy the global constraints
on the total magnetic flux produced by the dynamo and are able to
qualitatively reproduce the known statistical properties of the solar
cycle, like the Waldmeier's effects and the period - amplitude relation.
We expect that the model can be further developed taking into account
more accurately the turbulent properties of the sub-surface shear
layer. The accurate description of the magnetic helicity dissipation
is also important for the future progress. Direct numerical simulations
of the convective turbulence and helioseismological data analysis
techniques should help to improve our knowledge of the subsurface
shear layer and the physics of solar dynamo.
\section{Acknowledgements}

This work was supported by NASA LWS NNX09AJ85G grant and partially
by RFBR grant 10-02-00148-a.

\section{Appendix }

We describe some parts of the mean-electromotive force. The basic
formulation is given in P08. For this paper we reformulate tensor
$\alpha_{i,j}^{(H)}$, which represents the hydrodynamical part of
the $\alpha$-effect, by using Eq.(23) from P08 in the following form,
\begin{eqnarray}
\alpha_{ij}^{(H)} & = & \delta_{ij}\left\{ 3\eta_{T}\left(f_{10}^{(a)}\left(\mathbf{e}\cdot\boldsymbol{\Lambda}^{(\rho)}\right)+f_{11}^{(a)}\left(\mathbf{e}\cdot\boldsymbol{\Lambda}^{(u)}\right)\right)\right\} +\label{eq:alpha}\\
 & + & e_{i}e_{j}\left\{ 3\eta_{T}\left(f_{5}^{(a)}\left(\mathbf{e}\cdot\boldsymbol{\Lambda}^{(\rho)}\right)+f_{4}^{(a)}\left(\mathbf{e}\cdot\boldsymbol{\Lambda}^{(u)}\right)\right)\right\} \nonumber \\
 & + & 3\eta_{T}\left\{ \left(e_{i}\Lambda_{j}^{(\rho)}+e_{j}\Lambda_{i}^{(\rho)}\right)f_{6}^{(a)}+\left(e_{i}\Lambda_{j}^{(u)}+e_{j}\Lambda_{i}^{(u)}\right)f_{8}^{(a)}\right\} .\nonumber
\end{eqnarray}
 The contribution of magnetic helicity $\overline{\chi}=\overline{\mathbf{a\cdot}\mathbf{b}}$
($\mathbf{a}$ is a fluctuating vector magnetic field potential) to
the $\alpha$-effect is defined as $\alpha_{ij}^{(M)}=C_{ij}^{(\chi)}\overline{\chi}$,
where
\begin{equation}
C_{ij}^{(\chi)}=2f_{2}^{(a)}\delta_{ij}\frac{\tau_{c}}{\mu_{0}\overline{\rho}\ell^{2}}-2f_{1}^{(a)}e_{i}e_{j}\frac{\tau_{c}}{\mu_{0}\overline{\rho}\ell^{2}}.\label{alpM}
\end{equation}
 The turbulent pumping, $\gamma_{i,j}$, is also part of the mean
electromotive force in Eq.(23)(P08). Here we rewrite it in a more
traditional form (cf, e.g., \citealp{bra-sub:04,rad-kle-rog}),
\begin{equation}
\gamma_{ij}=3\eta_{T}\left\{ f_{3}^{(a)}\Lambda_{n}^{(\rho)}+f_{1}^{(a)}\left(\mathbf{e}\cdot\boldsymbol{\Lambda}^{(\rho)}\right)e_{n}\right\} \varepsilon_{inj}-3\eta_{T}f_{1}^{(a)}e_{j}\varepsilon_{inm}e_{n}\Lambda_{m}^{(\rho)},\label{eq:pump}
\end{equation}
 The effect of turbulent diffusivity, which is anisotropic due to
the Coriolis force, is given by:
\begin{equation}
\eta_{ijk}=3\eta_{T}\left\{ \left(2f_{1}^{(a)}-f_{1}^{(d)}\right)\varepsilon_{ijk}-2f_{1}^{(a)}e_{i}e_{n}\varepsilon_{njk}\right\} .\label{eq:diff}
\end{equation}
 Functions $f_{\{1-11\}}^{(a,d)}$ depend on the Coriolis number $\Omega^{*}=2\tau_{c}\Omega_{0}$
and the typical convective turnover time in the mixing-length approximation:
$\tau_{c}=\ell/u'$. They can be found in P08. The turbulent diffusivity
is parametrized in the form, $\eta_{T}=C_{\eta}\eta_{T}^{(0)}$, where
$\eta_{T}^{(0)}={\displaystyle \frac{u'\ell}{3}}$ is the characteristic
mixing-length turbulent diffusivity, $u'$ is the RMS convective velocity,
$\ell$ is the mixing length, $C_{\eta}$ is a constant to control
the intensity of turbulent mixing. The others quantities in Eqs.(\ref{eq:alpha},\ref{eq:pump},\ref{eq:diff})
are: $\mathbf{\boldsymbol{\Lambda}}^{(\rho)}=\boldsymbol{\nabla}\log\overline{\rho}$
is the density stratification scale, $\mathbf{\boldsymbol{\Lambda}}^{(u)}=\boldsymbol{\nabla}\log\left(\eta_{T}^{(0)}\right)$
is the scale of turbulent diffusivity, $\mathbf{e}=\boldsymbol{\Omega}/\left|\Omega\right|$
is a unit vector along the axis of rotation. Equations (\ref{eq:alpha},\ref{eq:pump},\ref{eq:diff})
take into account the influence of the fluctuating small-scale magnetic
fields, which can be present in the background turbulence and stem
from the small-scale dynamo (see discussions in \citealp{pouquet-al:1975a,moff:78,vain-kit:83,1996A&A...307..293K,2005PhR...417....1B}).
In our paper, the parameter $\varepsilon={\displaystyle \frac{\overline{\mathbf{b}^{2}}}{\mu_{0}\overline{\rho}\overline{\mathbf{u}^{2}}}}$,
which measures the ratio between the magnetic and kinetic energies
of fluctuations in the background turbulence, is assumed equal to
1. This corresponds to the energy equipartition. The quenching function
of the hydrodynamical part of $\alpha$-effect is defined by
\begin{equation}
\psi_{\alpha}=\frac{5}{128\beta^{4}}\left(16\beta^{2}-3-3\left(4\beta^{2}-1\right)\frac{\arctan\left(2\beta\right)}{2\beta}\right).
\end{equation}
 Note, in notation of P08 $\psi_{\alpha}=-3/4\phi_{6}^{(a)}$, and
$\beta={\displaystyle \frac{\left|\overline{B}\right|}{u'\sqrt{\mu_{0}\overline{\rho}}}}$.

\begin{figure}
\begin{center}
\includegraphics[width=0.9\textwidth]{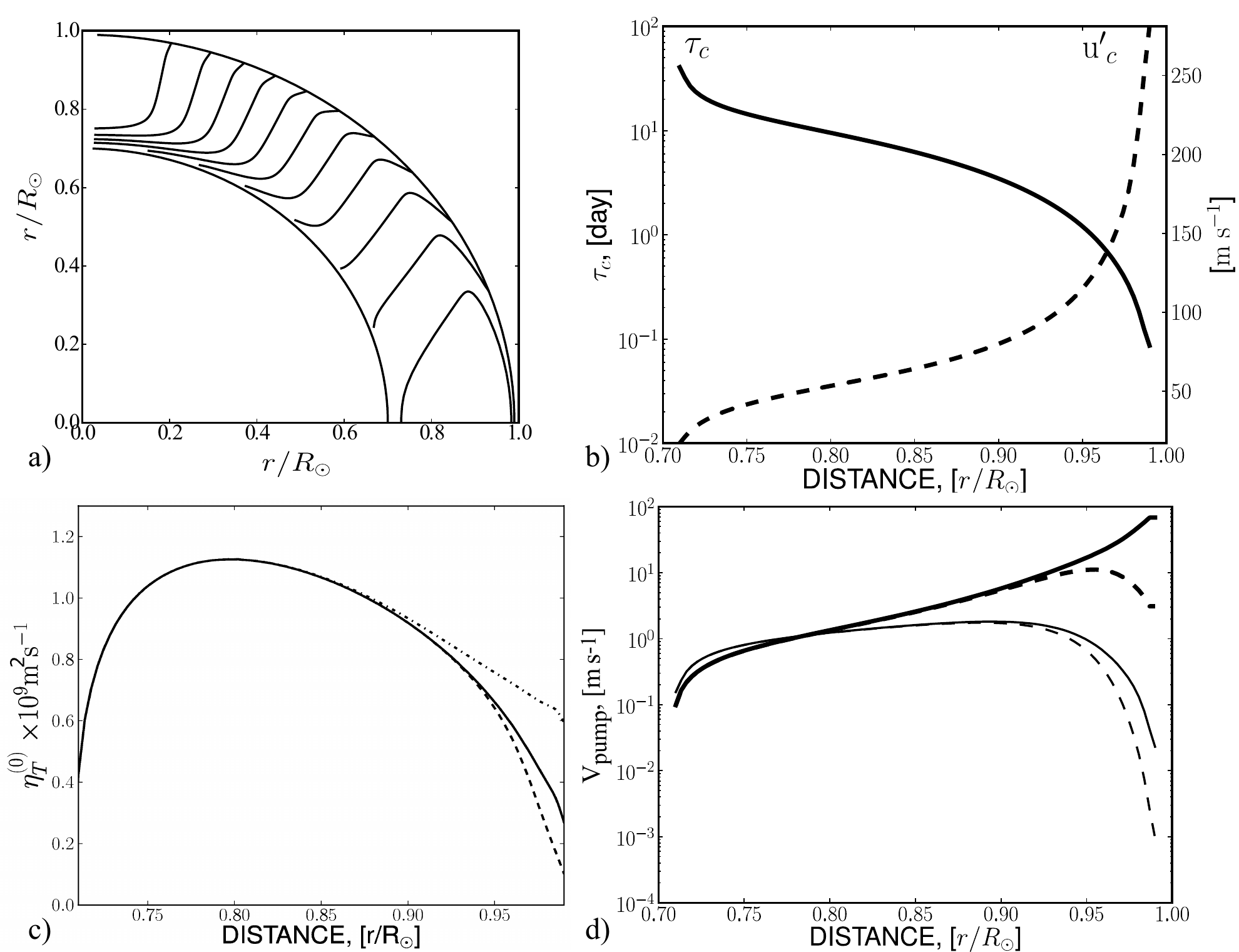}
\end{center}
\caption{\label{fig1-1}Parameters of the solar convection zone: a) contours
of the constant angular velocity plotted  in the range  $(0.75-1.05)\Omega_{0}$
with a step of $0.025\Omega_{0}$, $\Omega_{0}=2.86\cdot10^{-7}s^{-1}$;
b) turnover convection time $\tau_{c}$, and the RMS convective velocity
$u'_{c}$; c) the background turbulent diffusivity $\eta_{T}^{(0)}$
profiles; the solid curve shows the profile of the Stix(2002) model;
the dashed and dotted curves show models with the reduced and increased
sub-surface diffusivities discussed in Sec. 3.1; d) the radial (thick lines) and the latitudinal
(thin lines) turbulent pumping velocities at $\theta=45^{\circ}$ for
models P1 (solid curves) and P2 (dashed curves),  see Table 1;}
\end{figure}

\begin{figure}
\begin{center}
\includegraphics[width=\textwidth]{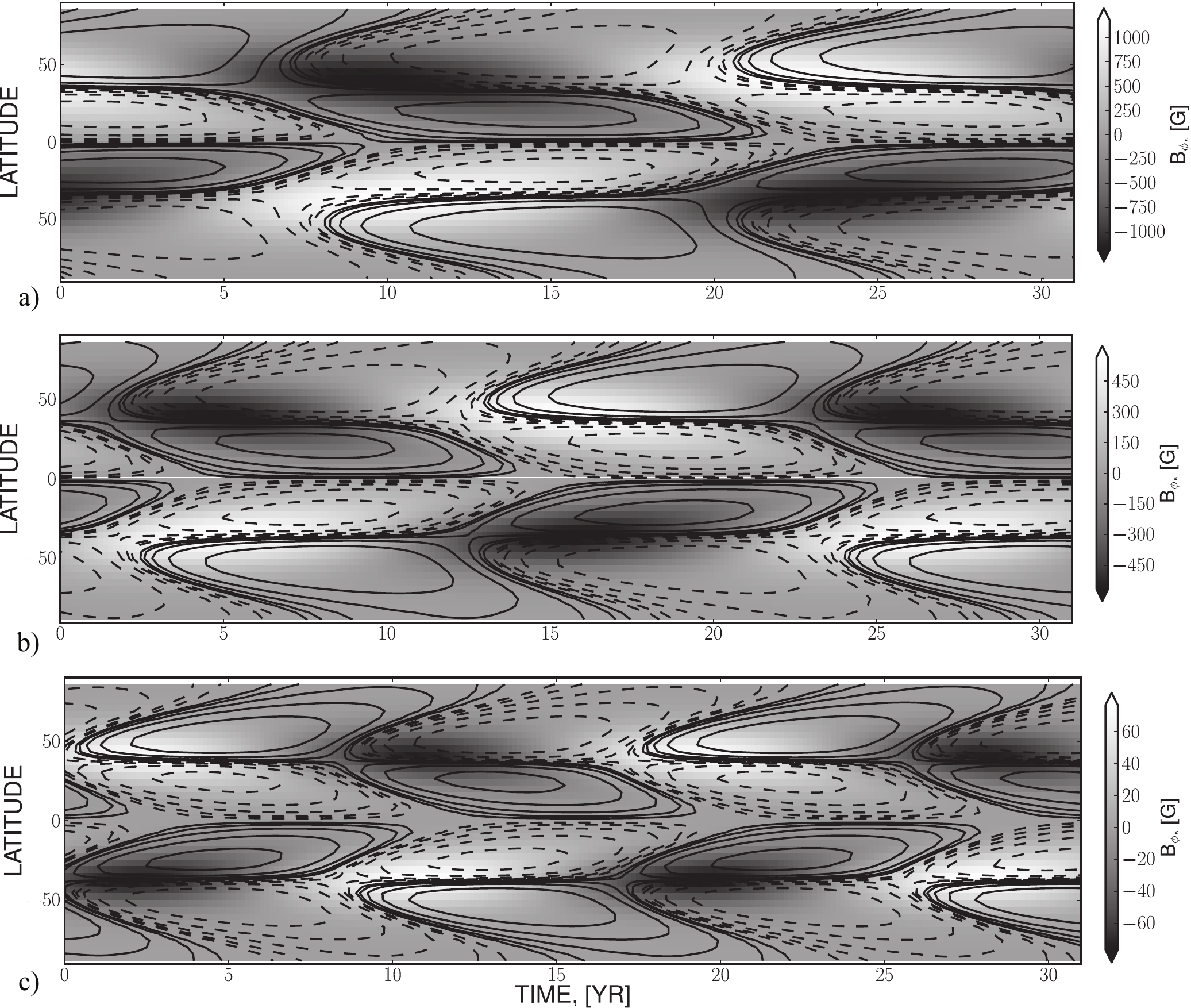}
\end{center}
\caption{Illustration of the influence of the magnetic pumping effect (models
P1, P2 and P3 in Table 1). Time-latitude diagrams of the toroidal magnetic
field averaged over the depth range $0.95-0.99R_{\odot}$ (gray scale)
and the radial field at $r=0.99R_{\odot}$(shown by contours), for
three cases of turbulent pumping: a) model P1 : the density gradient
profile provided by the \citet{stix:02} model, the radial field varies
in range $\pm21$G; b) model P2: the density gradient effect is restricted
as suggested by \citet{2000A&A...359..531K}, the radial field varies
in range $\pm14.5$ G; c) model P3: the pumping effect due to the
density gradient is neglected, the radial field varies in range $\pm1.4$G.}
\end{figure}

\begin{figure}
\begin{center}
\includegraphics[width=\textwidth]{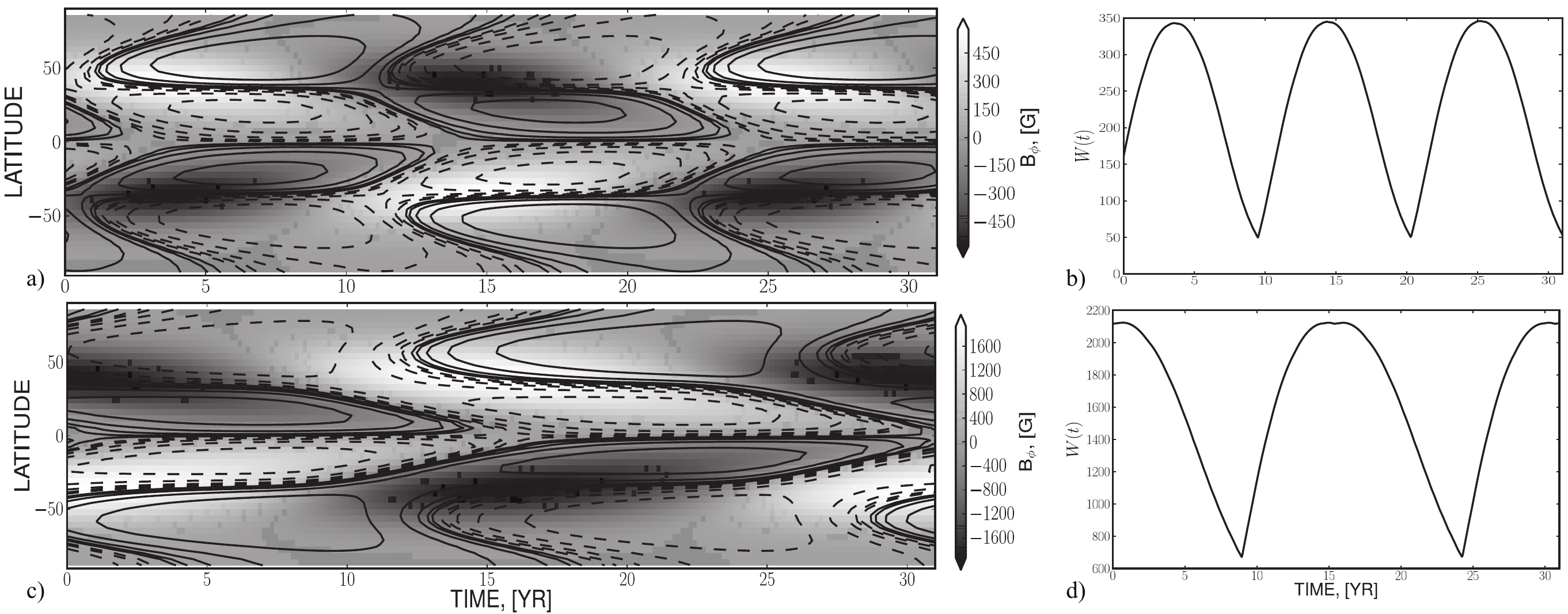}
\end{center}
\caption{Left column, the same as in Figure 2 for the larger (top) and smaller
(bottom) sub-surface turbulent diffusivity, models D1 and D2 respectively.
Right column - the simulated sunspot number, see definition in the
text.}
\end{figure}

\begin{figure}
\begin{center}
\includegraphics[width=\textwidth]{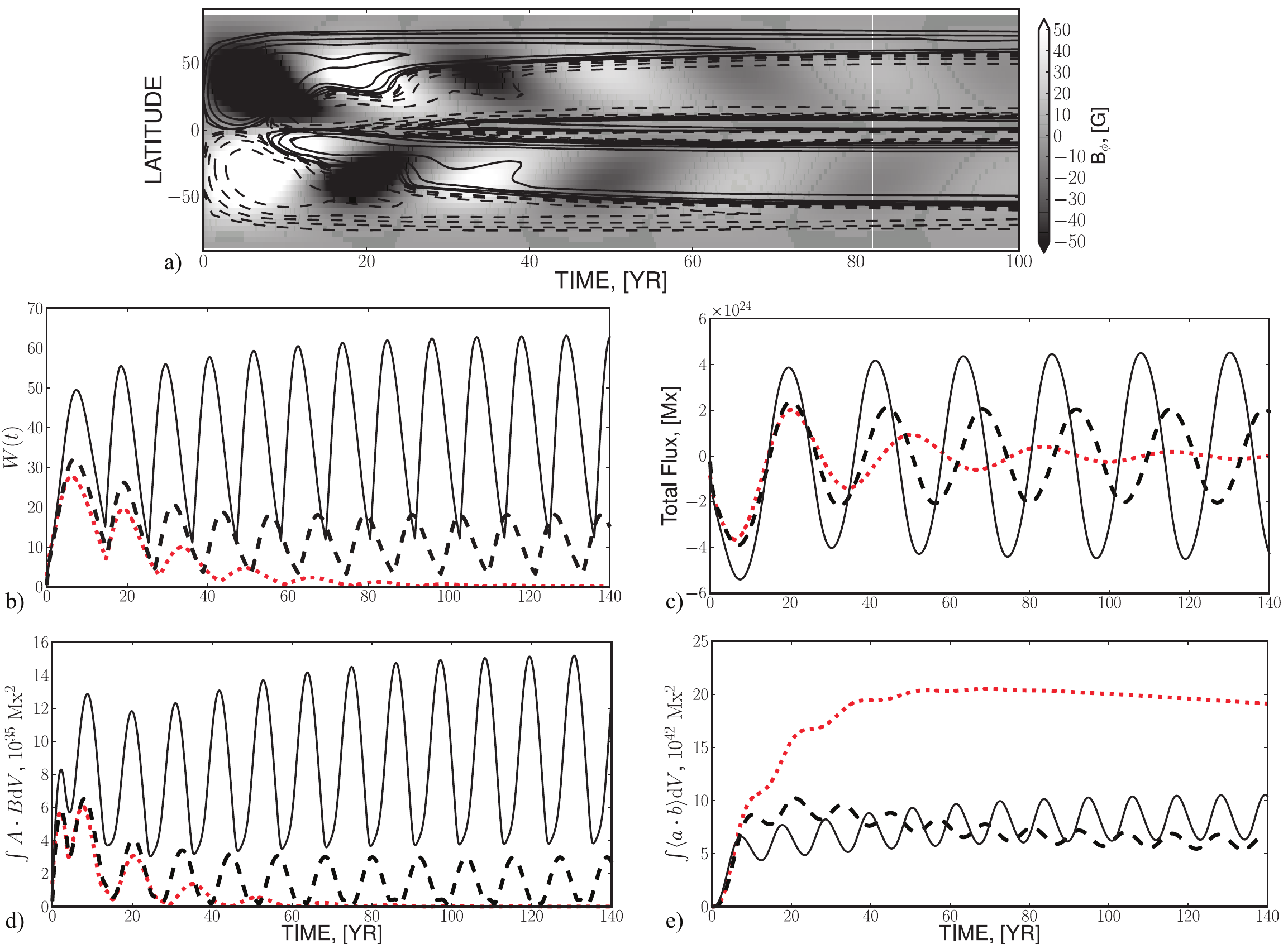}
\end{center}
\caption{Illustration of the dynamo models with the catastrophic quenching
(model CQ1) and without it (models CQ2 and CQ3): a) toroidal field
(gray scale)
 and current helicity (contours, in range  $\pm 3\ 10^{-5}$ G$^2$m$^{-1}$)
for model CQ1; evolution of global characteristics: b) the sunspot
number parameter W, c) the total magnetic flux, d) total large-scale
magnetic helicity; e) total small-scale magnetic helicity for models
CQ1(dots), CQ2(dashed curves), CQ3(solid curves).}
\end{figure}

\begin{figure}
\begin{center}
\includegraphics[width=0.6\textwidth]{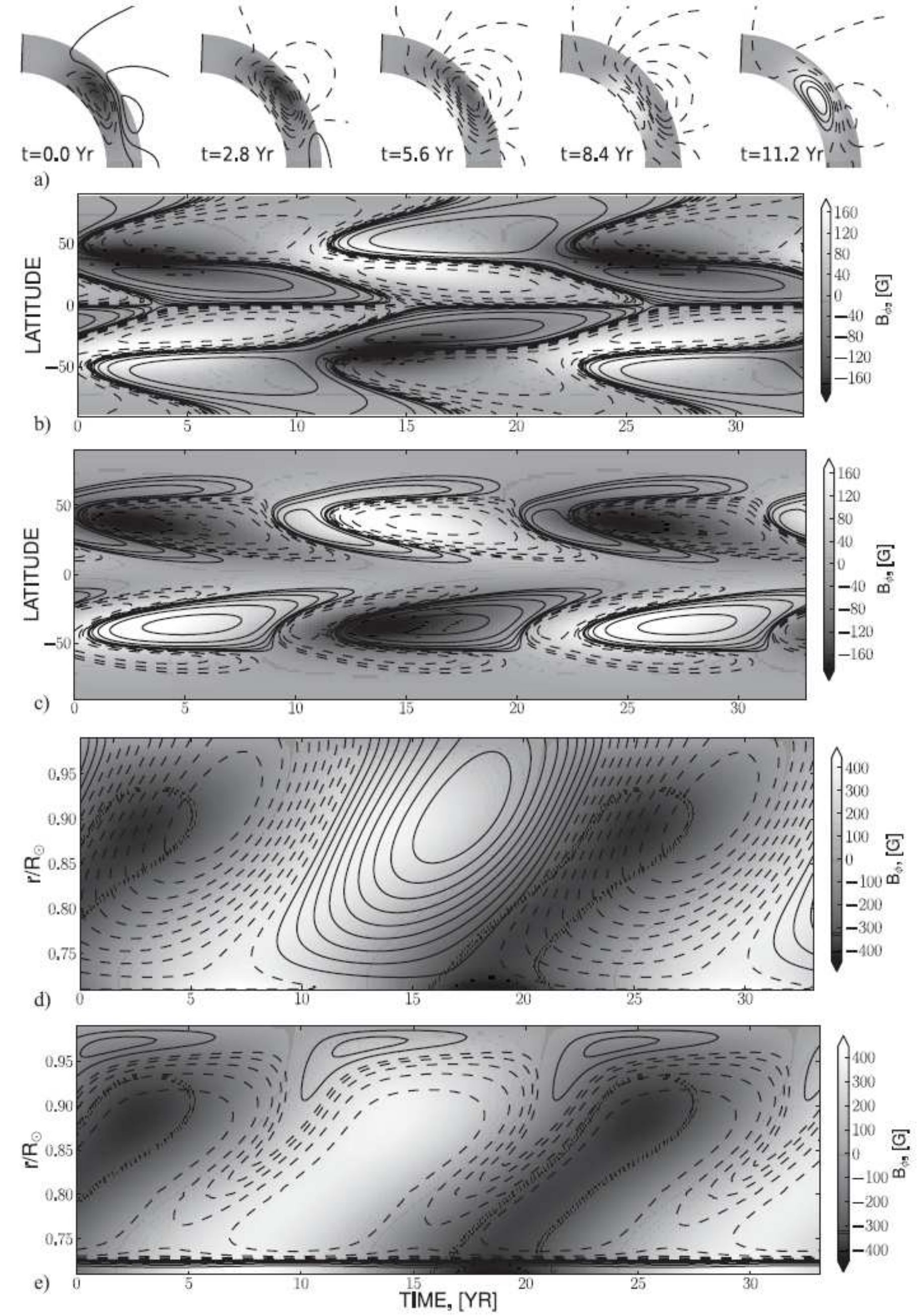}
\end{center}
\caption{Illustration of dynamo model CQ3 : a) snapshots of the toroidal
  (gray scale) and poloidal
(contours) magnetic field evolution for a half of the magnetic cycle;
 b) the time-latitude diagram for the
toroidal magnetic field averaged over the depth range
$0.95-0.99R_{\odot}$ (gray scale)  and the poloidal field (contours, in range $\pm3$ G);
 c) the time-latitude diagram for the
toroidal magnetic field (gray scale)  and the current
helicity (contours, in range $\pm 2\ 10^{-4}$ G$^2$m$^{-1}$);
 d) the time-radius  diagram at latitude $30^{\circ}$ for the
toroidal magnetic field (gray scale)  and the poloidal field (contours, in range $\pm3$ G);
e) the time-radius diagram at latitude  $30^{\circ}$ for the
toroidal magnetic field (gray scale)  and the current
helicity (contours, in range $-2\ 10^{-4}\div 2\ 10^{-3}$ G$^2$m$^{-1}$).}
\end{figure}

\begin{figure}
\begin{center}
\includegraphics[width=0.6\textwidth]{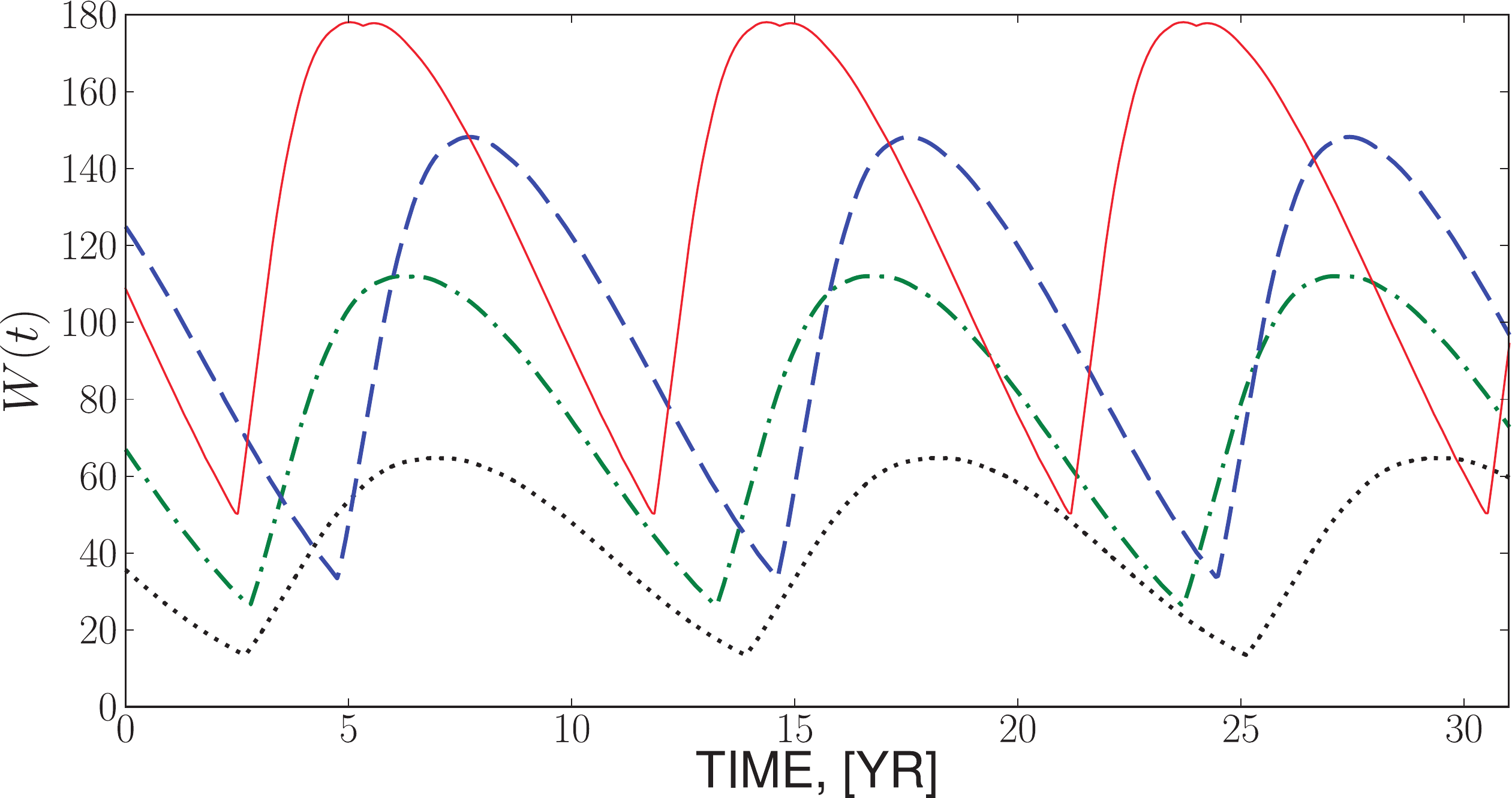}
\end{center}
\caption{The asymmetry of sunspot number parameter W in the series of dynamo
models WR2 (Table 1) calculated for various values of $C_{\alpha}$:
$0.03$(dots), $0.04$ (dot-dashed curve), $0.05$ (dashed curve) and
$0.06$ (solid curve).}
\end{figure}

\begin{figure}
\begin{center}
\includegraphics[width=0.9\textwidth]{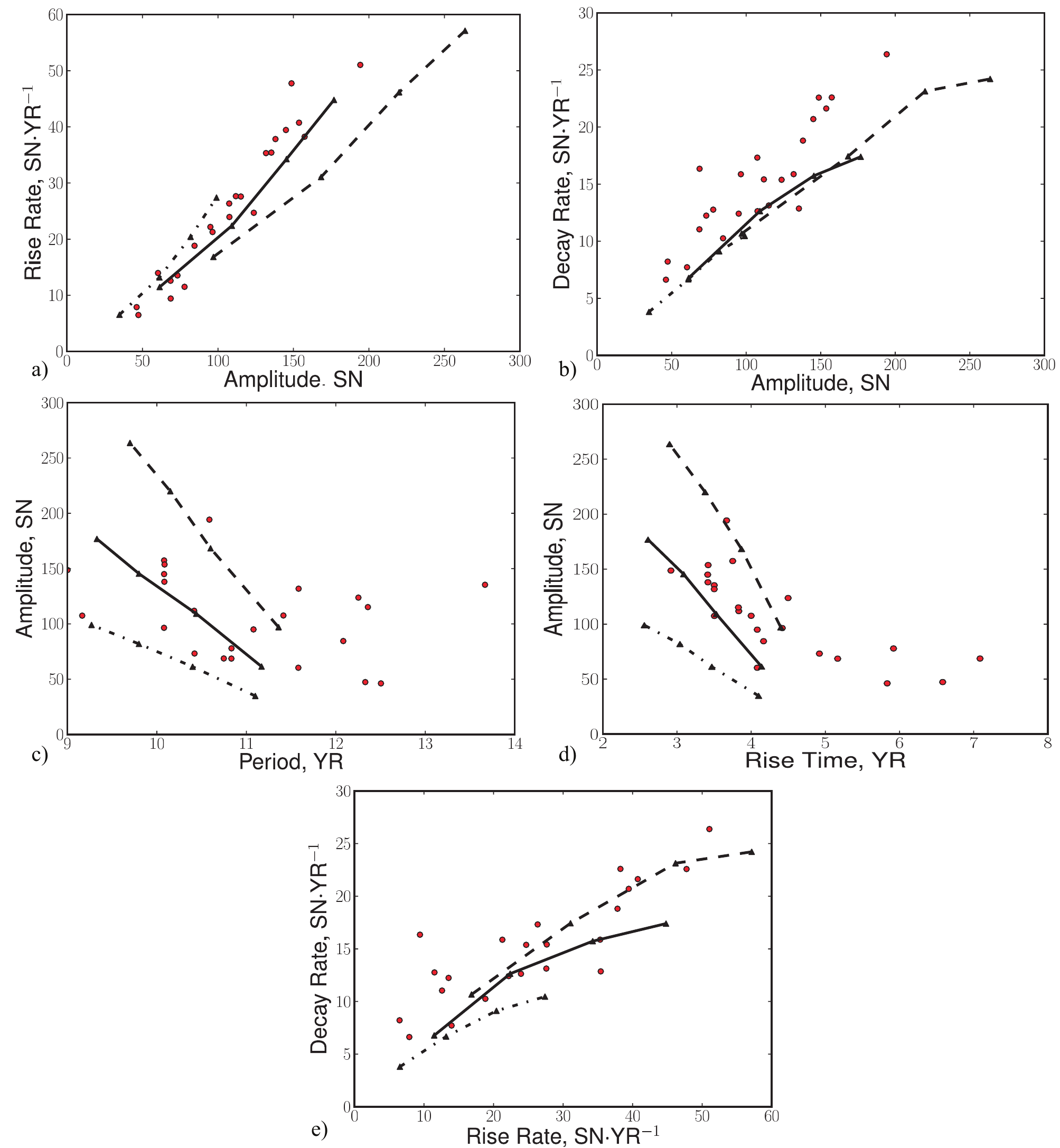}
\end{center}
\caption{The Waldmeier's (1935) relations in the models WR1 (dashed line),
WR2 (solid line) and WR3 (dot-dashed line): a) amplitude - rise rate,
b) amplitude - decay rate, c) period - amplitude, d) rise time - amplitude,
e) rise - decay rates, g) rise - decay times. The results from the
SIDC sunspot data are shown by circles.}
\end{figure}

\end{document}